\documentclass[11pt, epsfig]{article}
\usepackage{epsfig, amsmath, amssymb, amsthm, times}
\usepackage{graphicx}

\newcommand{\R}{{\mathbb R}}
\newcommand{\C}{{\mathbb C}}
\newcommand{\N}{{\mathbb N}}
\newcommand{\Z}{{\mathbb Z}}

\usepackage[mathcal]{eucal}

\newtheorem {thm}{Theorem}[section]

\theoremstyle{defintion}

\theoremstyle{remark}
\newtheorem{rem}[thm]{Remark}

\theoremstyle{example}

\theoremstyle{assumption}

\newcommand{\bnu}{\boldsymbol{\nu}}
\newcommand{\bphi}{\boldsymbol{\phi}}

\newcommand{\bW}{\boldsymbol{W}}
\newcommand{\bM}{\boldsymbol{M}}

\newcommand{\bT}{\boldsymbol{T}}
\newcommand{\bA}{\boldsymbol{A}}
\newcommand{\bR}{\boldsymbol{R}}
\newcommand{\bI}{\boldsymbol{I}}

\newcommand{\cF}{\mathcal{F}}

\newcommand{\cH}{\mathcal{H}}

\def\bI{\mathbf I}

\def\bR{\operatorname{\mathbf R}}
\def\bh{{\mathbf h}}
\def\T{\mathbb T}

\def\lbl{\label}
\def\be{\begin{equation}}
\def\ee{\end{equation}}
\def\p{\partial}

\newcommand{\1}{{i\mkern1mu}}

\title{ Instability of mixing in the Kuramoto model:
  From bifurcations to patterns}
\author{Hayato Chiba,\thanks{Advanced Institute for Materials Research,
    Tohoku  University, Sendai, 980-8557, Japan, {\tt hchiba@tohoku.ac.jp}}\;
Georgi S. Medvedev,\thanks{Department of Mathematics, 
Drexel University, 3141 Chestnut Street, Philadelphia, PA 19104,
{\tt medvedev@drexel.edu}} \;
and Matthew S. Mizuhara\thanks{Department of 
Mathematics and Statistics,
The College of New Jersey,
{\tt  mizuharm@tcnj.edu}}
}
\begin{document}
\maketitle
\begin{abstract}
We study patterns observed right after  the loss
of stability of mixing in the Kuramoto model of coupled phase oscillators with random
intrinsic frequencies on large graphs, which can also be random. We show that the emergent patterns
are formed via two independent mechanisms determined by the shape of the frequency distribution
and the limiting structure of the underlying graph
sequence. Specifically, we identify two nested eigenvalue
problems whose eigenvectors (unstable modes) determine the structure
of the nascent patterns. The analysis is
illustrated with the results of the numerical experiments with the
Kuramoto model with unimodal and bimodal
frequency distributions on certain graphs.
\end{abstract}

\section{Introduction}
\setcounter{equation}{0}
Models of interacting dynamical systems come up in different areas
of science and technology. Modern applications ranging from neuroscience to power grids 
emphasize models with spatially structured interactions defined by graphs.
Identifying dynamical mechanisms underlying pattern formation in such networks is an interesting
problem with many important applications. In this paper, we study patterns emerging near the loss
of stability of mixing in the  Kuramoto model (KM) with random
intrinsic frequencies on large graphs. We show that by varying  the frequency distribution
and the graph structure, one can generate a rich variety of spatiotemporal patterns
and identify a precise mathematical mechanism underlying pattern formation in this model.

The KM is one of the most widely used models in the theory of synchronization. 
In this paper, we study the KM on graphs, which is formulated as follows.
Let $(\Gamma^n)$ be a sequence
of graphs and consider
\begin{equation}\label{KM}
  \dot \theta_i=\omega_i +K(\alpha_nn)^{-1} \sum_{j=1}^n
  a^n_{ij}\sin(\theta_j-\theta_i),\quad i\in [n]:=\{1,2,\dots,n\},
\end{equation}
where $\theta_i:\R^+\to \T:=\R/2\pi\Z$ stands for the phase of oscillator $i$;
$\omega_i, i\in [n],$ are independent random intrinsic frequencies drawn from the distribution
with density $g(\omega),$ and $K$ is the strength of coupling. $(a^n_{ij})$ is an $n\times n$
symmetric (weighted) adjacency matrix of graph $\Gamma^n$. The scaling sequence $\alpha_n$
is either identically equal to $1$ if $(\Gamma^n)$ is a sequence of dense graphs, or
$\alpha_n\searrow 0$ subject to
condition $n\alpha_n\rightarrow \infty$ if $(\Gamma^n)$ is a sequence of sparse graphs
with edge density $O(n\alpha_n)$.
For more details on the KM on sparse graphs we refer an interested reader to
\cite{Med19}.

Suppose $(\Gamma^n)$ is a convergent sequence of  graphs whose
limiting behavior is described by a symmetric graphon $W\in L^2([0,1]^2)$ (cf.~\cite{Med19}).
Then under fairly general assumptions,
the dynamics of \eqref{KM} for large $n$ can be approximated by the Vlasov equation
\begin{equation}\label{MF}
\p_t f(t,\theta,\omega,x) +
\p_\theta \left\{ V(t,\theta,\omega,x) f(t,\theta,\omega,x) \right\}=0,
\end{equation}
where  $f(t,\theta,\omega,x)$ is a probability density function in $(\theta,\omega)\in \T\times\R$
of an oscillator at point $x\in I:=[0,1]$ at time $t\in \R^+$.
The velocity field is defined as follows
\be\lbl{def-V}
V(t,\theta,\omega,x)=\omega+ {K\over 2\1}  \left( \kappa(t,x) e^{-\1\theta} -\overline{\kappa(t,x)
    e^{-\1\theta}}\right),
\ee
where 
\be\lbl{corder}
\kappa(t,x)=\int_\T e^{i\theta} \int_\R  \left(\bW f(t,\theta,\omega,\cdot)\right)(x) d\omega d\theta,
\ee
is called a (local) order parameter. The following self-adjoint compact operator on $L^2([0,1])$ will play
an important role 
\be\lbl{def-W}
\left(\bW \phi \right)(x) :=\int_I W(x,y)\phi(y)dy, \quad I:=[0,1].
\ee

Rigorous justification of the Vlasov equation \eqref{MF} in the context of the KM with all--to--all
coupling was given in \cite{Lan05}. It is
based on the classical theory for kinetic equations (cf.~ \cite{Neu78, BraHep77,Dob79}).
For the KM on dense graphs, the Vlasov equation was further justified in \cite{KVMed18, ChiMed19a}.
For the KM on sparse graphs with unbounded degree, the results  in \cite{KVMed18, ChiMed19a}
continue to hold when combined with \cite[Theorem~4.1]{Med19}.

Equation \eqref{MF} has a steady state solution
\begin{equation}\label{steady}
  f_{mix}={g(\omega)\over 2\pi}.
\end{equation}
It describes the regime when all phases are uniformly distributed over $\T$, which corresponds
to mixing.
Numerical experiments show that mixing is stable for small $|K|$. In his classical work on
synchronization, Kuramoto identified the critical
value $K_c=2 \left(\pi g(0)\right)^{-1}$ marking the loss of stability of mixing \cite{Kur84}. This formula assumes
all--to--all coupling (i.e., $W\equiv 1$) and continuous even unimodal density $g(\omega)$.
Kuramoto's findings started a new area of research, which culminated in a rigorous analysis
of the loss of stability of mixing in the KM with all--to--all coupling in \cite{Chi15a}. 
For the KM on graphs, bifurcations of the mixing state were analyzed in \cite{ChiMed19a, ChiMed19b}.
Interestingly, the analysis of the spatially extended model along with the pitchfork bifurcation at
positive value of $K$ leading to synchronization reveals the possibility of a bifurcation for $K<0$.
For instance,
it was shown that for a network with nonlocal nearest-neighbor coupling there is a bifurcation to
so--called twisted states at a certain $K^-_c<0$ \cite{ChiMed19b, CMM18}. Thus, network structure plays
a role in the loss of stability of mixing and affects the emerging patterns.

From the beginning, the studies of the KM have been mainly focused on
the transition to synchronization, i.e., on the pitchfork bifurcation
of mixing. This turned out to be a challenging problem. The main technical obstacle to the bifurcation
analysis was the presence of continuous spectrum on the imaginary axis. It was overcome in
\cite{Chi15a} with the help of the generalized spectral theory. The method of \cite{Chi15a}
was further applied to the analysis of Andronov--Hopf bifurcation in \cite{Chi16} and was extended
to the KM on graphs in \cite{ChiMed19a, ChiMed19b}. The technical difficulty of the bifurcation
analysis for a long time obstructed its pattern formation aspect, which is perhaps even more
interesting from the nonlinear science point of view. The examples in \cite{CMM18}
already give
a glimpse into pattern formation capacity of spatially extended KM. In this note, we develop this theme
further. We show that the
combination of  frequency distribution and graph structure provides a flexible mechanism for
controlling spatiotemporal patterns arising at the loss of stability of mixing in the KM on graphs.
Surprisingly, the contributions of the frequency distribution and the graph structure are independent
from each other, which results in a variety of possible patterns obtained by combining features
controlled by these two (vector--valued) parameters (Figures~\ref{f.u-aa}-\ref{f.b-nn}).
Furthermore, we show that asymmetric frequency distribution results in asymmetric chimera like
patterns (Figure~\ref{f.asym-patterns}).
To keep the presentation simple, we restrict to linear stability analysis, which is sufficient to
relate bifurcations to patterns. Readers interested in the analysis beyond linear stability
are referred to \cite{ChiMed19b} for the treatment of the pitchfork bifurcation in the KM
on graphs. The Andronov--Hopf bifurcation is analyzed similarly by extending the results
in \cite{Chi16} to the spatially extended model following the lines of \cite{ChiMed19b}.

The outline of the paper is as follows. In the next section, we perform a linear stability analysis of
mixing. This is done for completeness, as the stability of mixing in the KM on graphs was already
analyzed in \cite{ChiMed19a, ChiMed19b}.  To complement the presentation in \cite{ChiMed19a}, this 
time we adapt the approach based on the theory for Volterra equations from \cite{Die16} to the
KM on graphs. It affords a quick derivation of the equation for the critical values of $K$ and
provides a nice geometric picture of the loss of stability of mixing in the KM.
After locating the instabilities, we compute the unstable modes,
which determine the bifurcating patterns. As shown in \cite{ChiMed19a} the loss of stability
in the KM on graphs is captured by two nested eigenvalue problems. The first one is obtained via the Fourier
transform of the linearized system and is the same as in the stability analysis of the KM with
all--to--all coupling \cite{Chi15a, Die16}. We refer to this problem as the
principal problem. The second one is that for $\bW$ (cf.~\eqref{def-W}) and is called a secondary
eigenvalue problem. It turns out that each problem is responsible for specific features of the bifurcating
solutions: the principal modes determine the local structure of the emerging
patterns, whereas the secondary modes capture their spatial organization.
In particular, the principal eigenvalue problem determines whether mixing loses stability via a pitchfork
or an Andronov--Hopf bifurcation. The former results in stationary patterns, whereas the latter produces
patterns traveling with a nonzero speed. However, it is the secondary eigenvalue problem that decides
the actual pattern. Depending on the form of the eigenfunctions corresponding to the critical eigenvalue
of the secondary problem, one can observe spatially uniform clusters or patterns with more complex
spatial organization like twisted states. Importantly, the two spectral problems are independent in the
sense that one is determined by the shape of the frequency distribution while the other - by the graph
structure. After
deriving the necessary analytical tools in Section~\ref{sec.linear}, we turn to the detailed discussion
of the bifurcating patterns in the KM with unimodal and bimodal $g$ in Section~\ref{sec.patterns}.
To this end, we compare solutions bifurcating from the mixing state in the KM on all--to--all and
nonlocal nearest--neighbor graphs. These examples clearly demonstrate the contributions of the
principal and secondary unstable modes to the structure of the emerging patterns.

\section{Linear stability}\label{sec.linear}
\setcounter{equation}{0}

In this section, we rewrite \eqref{MF} in Fourier variables and linearize the resultant system about the
mixing steady state. Then we reduce the problem of stability to the Volterra equation for vector--valued
functions and derive the equation for the eigenvalues of the linearized operator.
Here, we extend the method in \cite{Die16} to the KM on graphs. Then we compute the corresponding
eigenvalues following \cite{ChiMed19b}. This information is sufficient to explain the patterns emerging
at the loss of stability of mixing, which is the main focus of this paper. For more details on stability
analysis of mixing in the KM on graphs, we refer an interested reader to \cite{ChiMed19a, ChiMed19b}.

We start by applying the Fourier transform in $(\theta,\omega)$
\be\lbl{2FT}
u(t,l,\xi, x)= \int_\T\int_\R  e^{\1l\theta} e^{\1\xi\omega} f(t,\theta, \omega, x) d\omega d\theta,
\quad (l,\xi, x)\in \Z\times\R\times I
\ee
to \eqref{MF}. 
Note that by the definition of $f,$
\be\lbl{marginal}
g(\omega)=\int_\T f(t,\theta,\omega,x) d\theta\quad \forall (x,t)\in I\times\R^+.
\ee
Thus, by Fubini's theorem,
\be\lbl{def-hat}
u(t,0,\xi,x)=\int_\R  e^{\1\xi\omega} g(\omega) d\omega=: (\cF g)(\xi),
\ee
where $\cF g$ stands for the Fourier transform in $\omega$
throughout this paper.

Following \cite{Die16}, we assume
\begin{equation}\label{decay}
\cF g\in C_a(\R)=\{ \phi\in C(\R):\; \|\phi\|_a=\sup_{t\in\R} e^{at} |\phi(t)|<\infty \}\; \mbox{for some}\; a>0.
\end{equation}

Further, since $f$ is real, $\overline{u(t,l,\xi,x)}=u (t,-l,-\xi,x)$ it is sufficient to
consider\footnote{Note that $\p_t u(t,0,\xi, x)=0,\;
(\xi, x)\in\R\times I$ by \eqref{def-hat}.}
\begin{eqnarray}\lbl{FT1}
\p_t u(t,1,\xi,x) &=& \p_\xi u(t,1,\xi,x) + {K\over 2}\left(\kappa(t,x) (\cF g)(\xi) -\overline{\kappa(t,x)} u(t,2,\xi,x)\right),\\
\lbl{FT2}
  \p_t u(t,l,\xi,x) &=& l \p_\xi u(t,l,\xi,x) + {Kl\over 2}\left( \kappa(t,x) u(t, l-1, \xi, x) -
                        \overline{\kappa(t,x)} u(t,l+1,\xi,x)\right), 
\end{eqnarray}
for $\ge 2$.  Note that $\kappa$ defined in \eqref{corder} can be rewritten as
\begin{equation}\label{rewrite-kappa}
  \kappa(t,x)=\int_I W(x,y) u(t,1,0,y)dy.
  \end{equation}
  
The equilibrium $f_{mix}$ corresponds to 
$u_{mix}=(\mathcal{F} g, 0,0,\dots)$ in the Fourier space for $l\in \{0,1,2,\dots\}$. 
The linearization of about $u_{mix}$ is thus given by  
\begin{eqnarray}\lbl{L1}
\p_t u(t,1,\xi,x) &=& \p_\xi u(t,1,\xi,x) + {K\over 2}\kappa(t,x) (\cF g)(\xi),\\
\lbl{L2}
\p_t u(t,l,\xi,x) &=& l \p_\xi u(t,l,\xi,x), \; l\ge 2.
\end{eqnarray}
Equations in \eqref{L2} describe pure transport. Thus, stability is decided by \eqref{L1},
which we rewrite as
\begin{equation}\label{linear-operator}
  \p_t\phi =\bT\phi.
\end{equation}
$\bT$ is viewed as a linear operator densely defined on $L^2\left([0,1]; C_a(\R)\right).$

Integrating \eqref{L1} along characteristics and recalling \eqref{rewrite-kappa}, we have
\begin{equation}\lbl{ch1}
  u(t,1,\xi,x) = u_0(1,\xi+t,x) +{K\over 2} \int_0^t \left(\bW u(s,1,0,\cdot)\right) (\cF g) (\xi+(t-s)) ds,
\end{equation}
By plugging $\xi=0$ in \eqref{ch1}, we obtain the following Volterra equation
\begin{equation}\lbl{Volt}
  u(t,x)= u_0(t,x)+{K\over 2} \int_0^t \left(\bW u(s,\cdot)\right) (\cF g) (t-s) ds,
\end{equation}
where by abuse of notation $u(t,x):=u(t,1,0,x)$.
We recast \eqref{Volt} in a more general form
 \begin{equation}\label{infV}
         \bnu(t) +\int_0^t \bA(t-s) \bnu(s) ds = \bphi(t),
         \end{equation}
         where $\bA:\R^+\to L(\cH)$ and $\bphi, \bnu:\R\to \cH$.
By $L(\cH)$ we denote the space of linear bounded operators on $\cH$. 
         For the problem at hand, $\cH=L^2([0,1])$,
         \begin{equation}\label{specify-A}
           \bA(t)={-K \over 2} (\cF g)(t)\bW\quad\mbox{and}\quad \bphi(t)=u_0(t,\cdot),\; \bnu(t)=u(t,\cdot).
         \end{equation}
         From now on, we will use the bold font to denote  vector--valued functions along with operators.

         \begin{thm}\label{thm.resolvent}\cite[Theorems~1 \& 2]{Grip87}
           Let $\bA:\R^+\to L(\cH)$ be strongly measurable and $\|\bA(\cdot)\|\in L^1_{loc}(\R^+;\R),$
           where $\|\cdot\|$ stands for the operator norm.
         Then there exists a strongly measurable resolvent $\bR:\R^+\to L(\cH)$
         \begin{equation}\label{abstract-resolvent}
           \bR(t)=\bA(t)-\int_0^t \bR(t-s) \bA(s) ds= \bA(t)-\int_0^t \bA(t-s)\bR(s)ds.
         \end{equation}
         For any $\bphi\in L^1_{loc}(\R^+,\cH)$ the unique solution of \eqref{infV}
         can be expressed as
         \begin{equation}\label{solve-infV}
           \bnu(t)=\bphi(t)-\int_0^t \bR(t-s)\bphi (s)ds.
         \end{equation}
         Moreover, $\bR\in L^1(\R^+; L(\cH))$ if and only if
                  $\bI +(\mathcal{L}\bA) (z)$ is invertible
         as a bounded operator on $\cH$ for all $z\in\C$ with $\Re z\ge 0$. Here,
         \begin{equation}\label{LapA}
           \mathcal{L}\bA (z)=\int_0^t e^{-zt} \bA (t) dt.
         \end{equation}
       \end{thm}
       \begin{rem}\label{rem.explicit} For \eqref{Volt}
following the lines of the analysis in the finite--dimensional case \cite[Theorem~3.1]{GLS90},
         the resolvent can be obtained  constructively as a  Neumann
         series 
         $$
\bR\bh = -\sum_{j=1}^\infty {k^\ast}^j\ast (\bW^j \mathbf{h}),\quad k(t)={K \over 2} (\cF g)(t),
$$
or in expanded form
$$
(\bR h)(t,x)= -\sum_{j=1}^\infty \int_0^t {(k^\ast}^j (t-s) \left(\bW^j h(s,\cdot)\right)(x) ds.
$$
Here, $k\ast a$ stands for the convolution of two functions $(k\ast a)(t)=\int_0^t k(t-s)a(s) ds$.
Similarly,
$$
{k^\ast}^j\ast a= \underbrace{
    k\ast (k\ast\dots\ast ( k\ast
      }_{j\,\mbox{times} }
   a )\dots).
$$
\end{rem}

       The data in \eqref{Volt} clearly satisfy the
       assumptions of Theorem~\ref{thm.resolvent}.
       Our next goal is to understand invert-ability of
       $\bM(z)=\bI+(\mathcal{L}\bA)(z)$. By \eqref{specify-A},
         $\bM(z)$ is invertible unless $z$ is a root of
           \begin{equation}\label{spectral}
             G(z)={2\over K\mu},\; G(z):=\mathcal{L} \left(\cF g\right)(z),\quad
             \mu\in\operatorname{Spec}(\bW)/\{0\}.
           \end{equation}
           Below, we will need the following observation
           \be\lbl{compute-Lapk}          
G(z) = \int_0^\infty (\cF g)(t) e^{-tz} dt = \int_0^\infty\int_{-\infty}^\infty
e^{\1 (\xi+t)\eta} e^{-tz}  g(\eta) d\eta dt 
=  \int_{-\infty}^\infty {g(\eta)\over z-\1\eta} d\eta,
\ee
where we used Fubini's theorem.

          Equation \eqref{spectral} will be used to compute the eigenvalues of $\bT$.
           The corresponding eigenfunctions can be found by extending the corresponding results
           of the theory for Volterra equations on a finite--dimensional space
           (cf.~\cite[Theorem~2.1]{GLS90}) to the problem at hand. For simplicity of presentation,
           we will compute the eigenfunctions directly using the results in \cite{ChiMed19a}.

           Let $z=\lambda$ be a root of \eqref{spectral} corresponding to
           $\mu\in\operatorname{Spec}(\bW)$. By Lemma~3.2 in \cite{ChiMed19a}
           (see also \cite[Lemma~27]{Die16}), $\lambda$ is an eigenvalue of $\bT$. The corresponding
           eigenfunction is
           \be\lbl{Fv}
           w_\lambda=\mathcal{F} v_\lambda,
           \ee
           where
           \begin{equation}\label{v-through-V}
             v_\lambda (\omega, x)={K\over 2} {g(\omega)\over \lambda -\1\omega}
             \int_{I\times\R} W(x,y) v(\lambda,y) dyd\lambda.
           \end{equation}

           By integrating both sides of \eqref{v-through-V} with respect to $\omega,$ we
           have
           $$
           V={K\over 2} G(\lambda) \bW V,
           $$
           where we used \eqref{compute-Lapk}.
           $V(x)=\int_\R v(\lambda, x)d\lambda$ is an eigenfunction of $\bW$ corresponding to
           $$
           \mu={K\over 2} G(\lambda).
           $$
           We conclude that
           \begin{equation}\label{compute-w}
             w_\lambda(\eta,x)=\int_\R {e^{\1\eta\omega} g(\omega)\over \lambda -\1\omega} d\omega\; V(x)
           \end{equation}
           is an eigenfunction of $\bT$ corresponding to eigenvalue $\lambda$.
           Here, we dropped the factor $K\mu/2$ since eigenfunctions are defined up to a multiplicative
           constant.

           Thus,
           \be\lbl{v-lambda}
           v_\lambda(\omega,x)=\left(\mathcal{F}^{-1}w_\lambda(\cdot,x)\right)(\omega)=
           \Upsilon_\lambda(\omega) V(x), \quad \Upsilon_\lambda (\omega):=
           {g(\omega)\over \lambda-\1\omega}.
           \ee
           \begin{rem}
             \lbl{rem.sep}
           The separable structure of $v_\lambda$ has important implications for pattern formation.
           $\Upsilon$ and $V$ are determined by the intrinsic frequency distribution $g$  and the graph limit $W$
           respectively. Equation~\eqref{v-lambda} shows that the frequency distribution and the graph structure
           shape the unstable modes independently from each other.
         \end{rem}
         
           Below we will need to know the structure of
           $$
            \Upsilon_0=\lim_{\lambda \to 0+0}  \Upsilon_\lambda = \lim_{\lambda \to 0+0} {g(\omega)\over \lambda-\1\omega}.
           $$
           For $(\cF g)\in C_a(\R)$, $\Upsilon_0$ can be viewed as a tempered distribution. Indeed, for any $\phi$ from the Schwartz
           space $\mathcal{S}(\R)$, by Sokhotski--Plemelj formula (cf.~\cite{Simon-Complex}), we have
           $$
           \langle  \Upsilon_0,\phi\rangle =\lim_{\lambda \to 0+0} \int_{-\infty}^\infty {g(\omega)\phi(\omega)\over \lambda-\1\omega}d\omega
           =\pi g(0)-\1\operatorname{P.V.} \int_{-\infty}^\infty {g(\omega)\over \omega} d\omega.
           $$
           Thus, as an element of $\mathcal{S}^\prime (\R)$, $ \Upsilon_0$ can be written as
           \be\lbl{v0-distribution}
            \Upsilon_0=\pi g(0) \delta-\1\mathcal{P}_{g,0},
           \ee
           where $\delta$ stands for the delta function and
           $$
           \langle \mathcal{P}_{g,\alpha}, \phi \rangle=\operatorname{P.V.}
             \int_{-\infty}^\infty {g(\omega+\alpha)\phi(\omega+\alpha)\over \omega}d\omega.
           $$

           Similarly, we compute
           \be\lbl{viy-distribution}
             \Upsilon_{\pm\1 y^\ast}:=\lim_{\lambda (0+0) \pm\1y^\ast } {g(\omega)\over \lambda-\1\omega}=
           \pi g(\pm y^\ast) \delta_{\pm y^\ast}-\1\mathcal{P}_{g, \pm y^\ast},
           \ee
where $\delta_{\beta}=\delta(\cdot+\beta).$
                      
\section{Bifurcations and patterns}\label{sec.patterns}
\setcounter{equation}{0}

Next we turn to the bifurcations in the KM \eqref{KM} and the corresponding patterns.
To illustrate the typical scenarios realized in this model we will consider unimodal
(\textbf{U}) and bimodal (\textbf{B}) $g\in C_a(\R)$ combined with all--to-all (\textbf{aa})
and nonlocal nearest-neighbor connectivity (\textbf{nn}). We will code the corresponding
models by (\textbf{Xy}) where $\mbox{\bf X}\in \{\mbox{\bf U},\mbox{\bf B}\}$
and $\mbox{\bf y}\in \{\mbox{\bf aa},\mbox{\bf nn}\}$.
\begin{figure}
	\centering
\textbf{a}\;	\includegraphics[width=.3\textwidth]{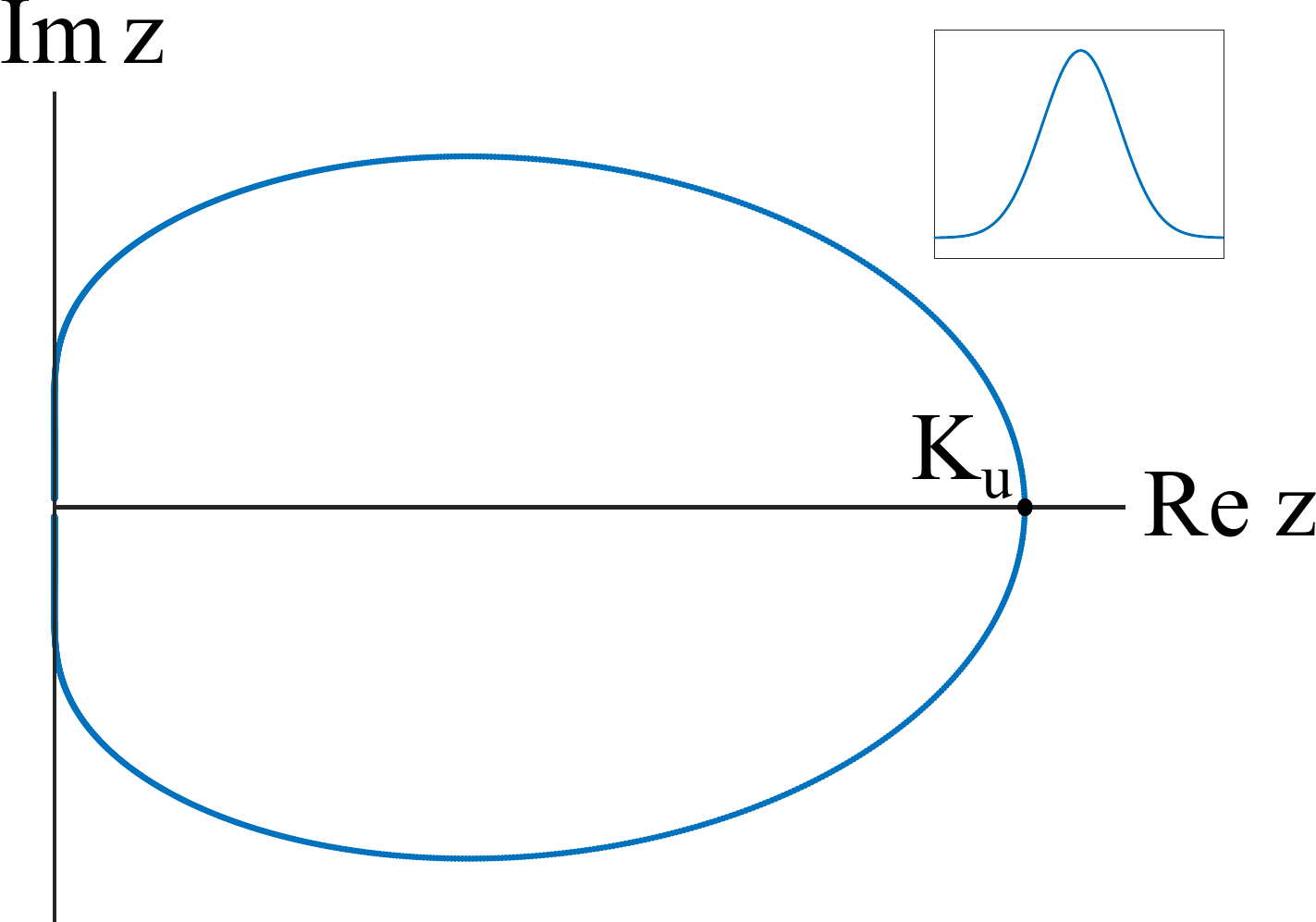}\quad\quad\quad
 \textbf{b}\;       \includegraphics[width=.3\textwidth]{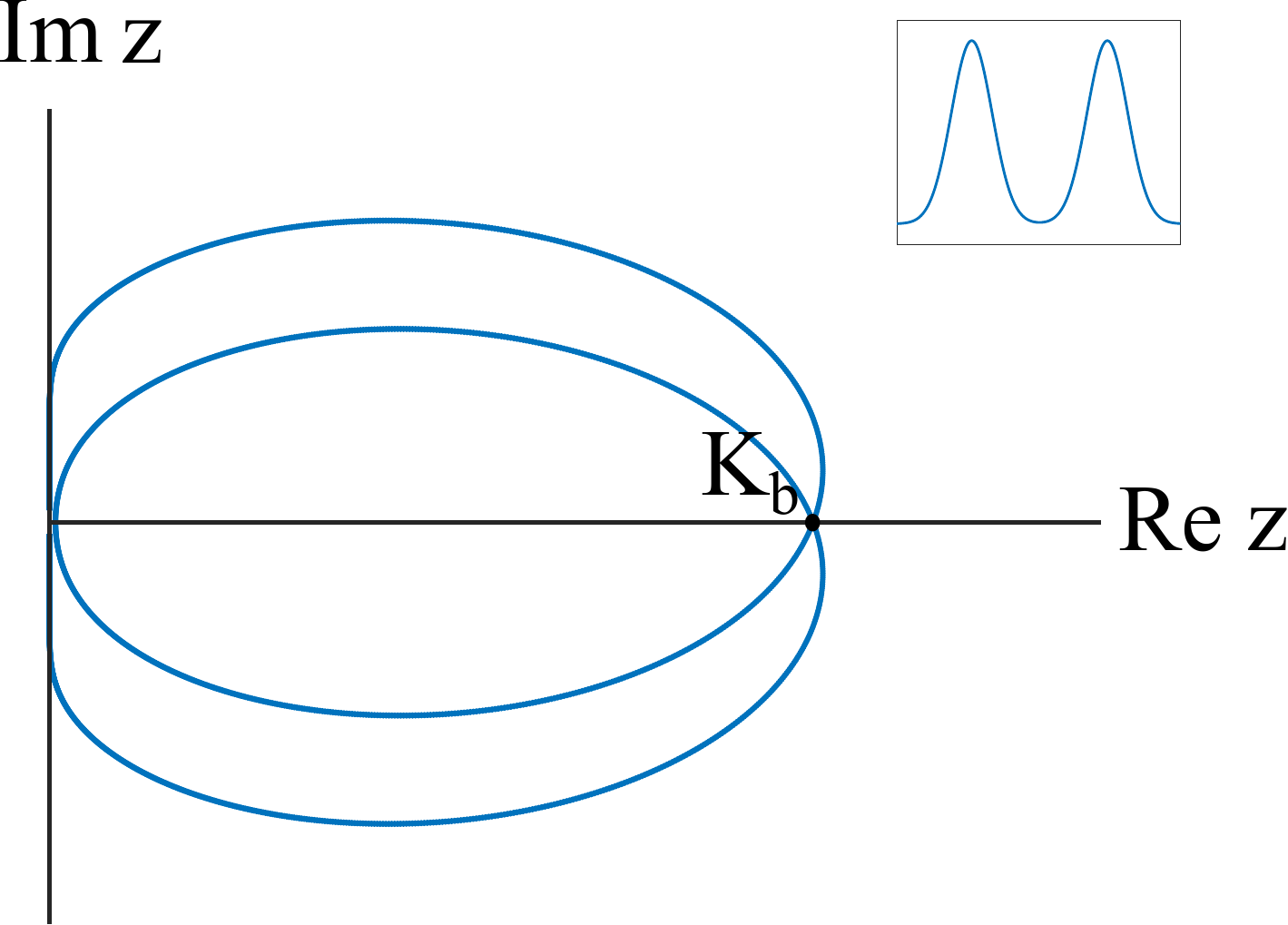}
 \caption{The critical curves generated by a unimodal (\textbf{a}) and bimodal (\textbf{b})
   intrinsic frequency distribution.}
	\label{f.curve}
\end{figure}

As a  first step, we  locate the bifurcations in \eqref{KM} by solving \eqref{spectral}.
To this end, recall \eqref{compute-Lapk} and note that
$G$ is analytic in $\Pi:=\{z\in\C:\; \Re z>0\}$ and is continuous in
$\bar\Pi.$
By Sokhotski--Plemelj formula,
\be\lbl{ytoinfty}
\lim_{y\to\pm\infty}\lim_{x\to 0+0} G(x+\1 y)=\lim_{y\to\pm\infty} \left( \pi g(y)-\1
  \operatorname{P.V.} \int_{-\infty}^\infty {g(s)ds \over y-s}\right)=0.
\ee

Thus, $G$ maps the imaginary axis to a bounded closed curve.
As in \cite{Die16}, we use the Argument Principle \cite{Simon-Complex}
to conclude that \eqref{spectral}
has a root in $\Pi$ if and only if
\begin{equation}\label{Penrose}
  2(K\mu)^{-1}\in G(\Pi).
\end{equation}
\begin{figure}
	\centering
	\includegraphics[width =.6\textwidth]{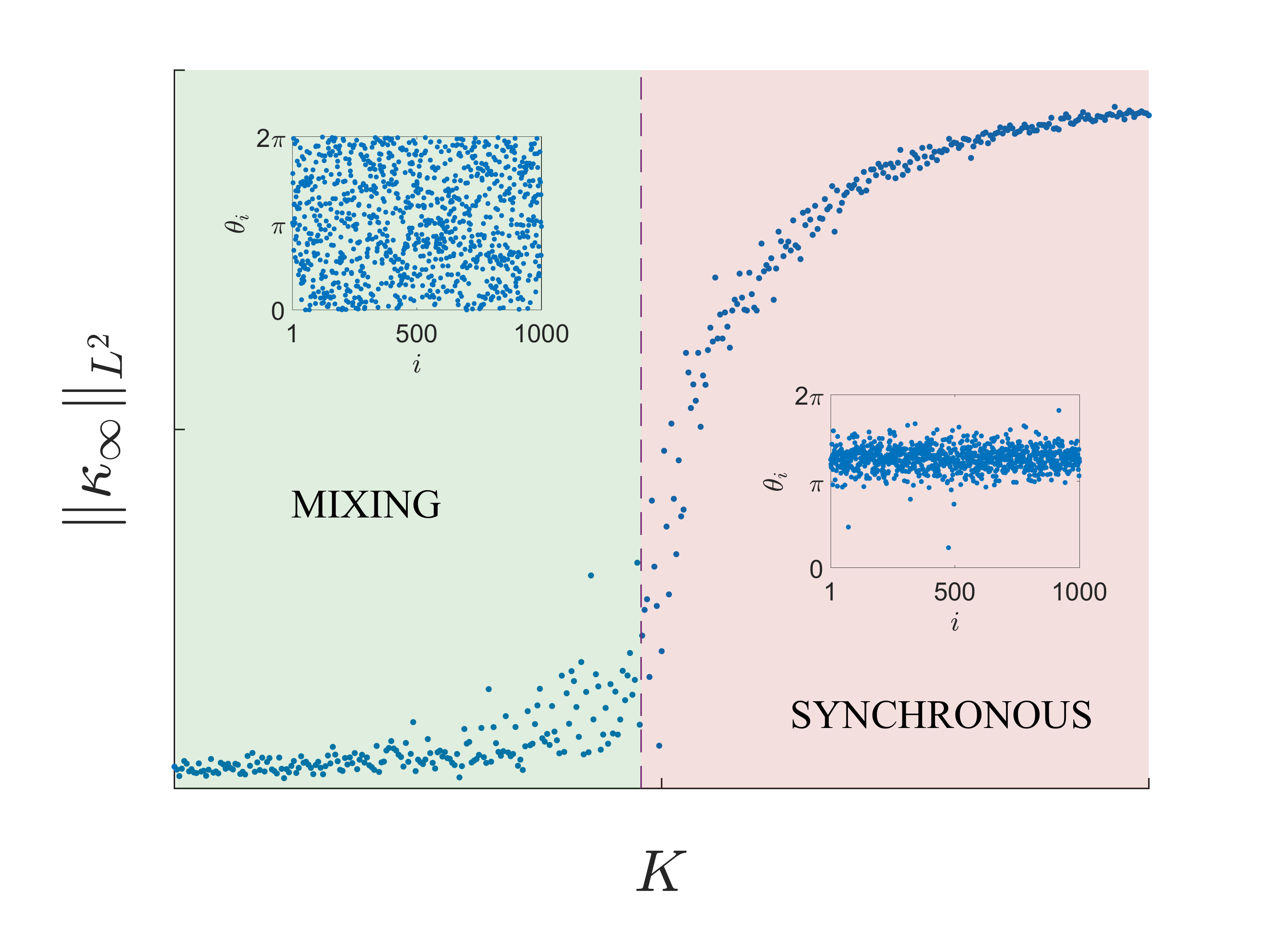}
	\caption{The pitchfork bifurcation leading to the formation of spatially homogeneous
          synchronous solution in the KM with all--to--all coupling and unimodal $g$.
          $\kappa_\infty$ stands for the time asymptotic value of the $L^2$--norm  of the local
          order parameter \eqref{corder}. The data presented in this plot was obtained after a
          long transient time. Further, the $L^2$--norm of the local order parameter was averaged
          over a certain time interval to smoothen the data. The same method was used in all
          subsequent computations of the order parameter.
        }
	\label{f.u-aa}
\end{figure}

In Figure~\ref{f.curve}, we plot $G(\p\Pi)$ for symmetric unimodal and bimodal $g$. 
The critical curve $G(\p\Pi)$  always intersects the real axis
at the origin (cf.~\eqref{ytoinfty}). In addition, it has another point
of intersection with the real axis at
\be\lbl{Ku}
K_u=G(0)>0,
\ee
if $g$ is unimodal. In the bimodal case, there is a point of double intersection
\be\lbl{Kb}
K_b=G(\pm\1 y^\ast)>0.
\ee

\begin{figure}
	\centering
	\includegraphics[width =.6\textwidth]{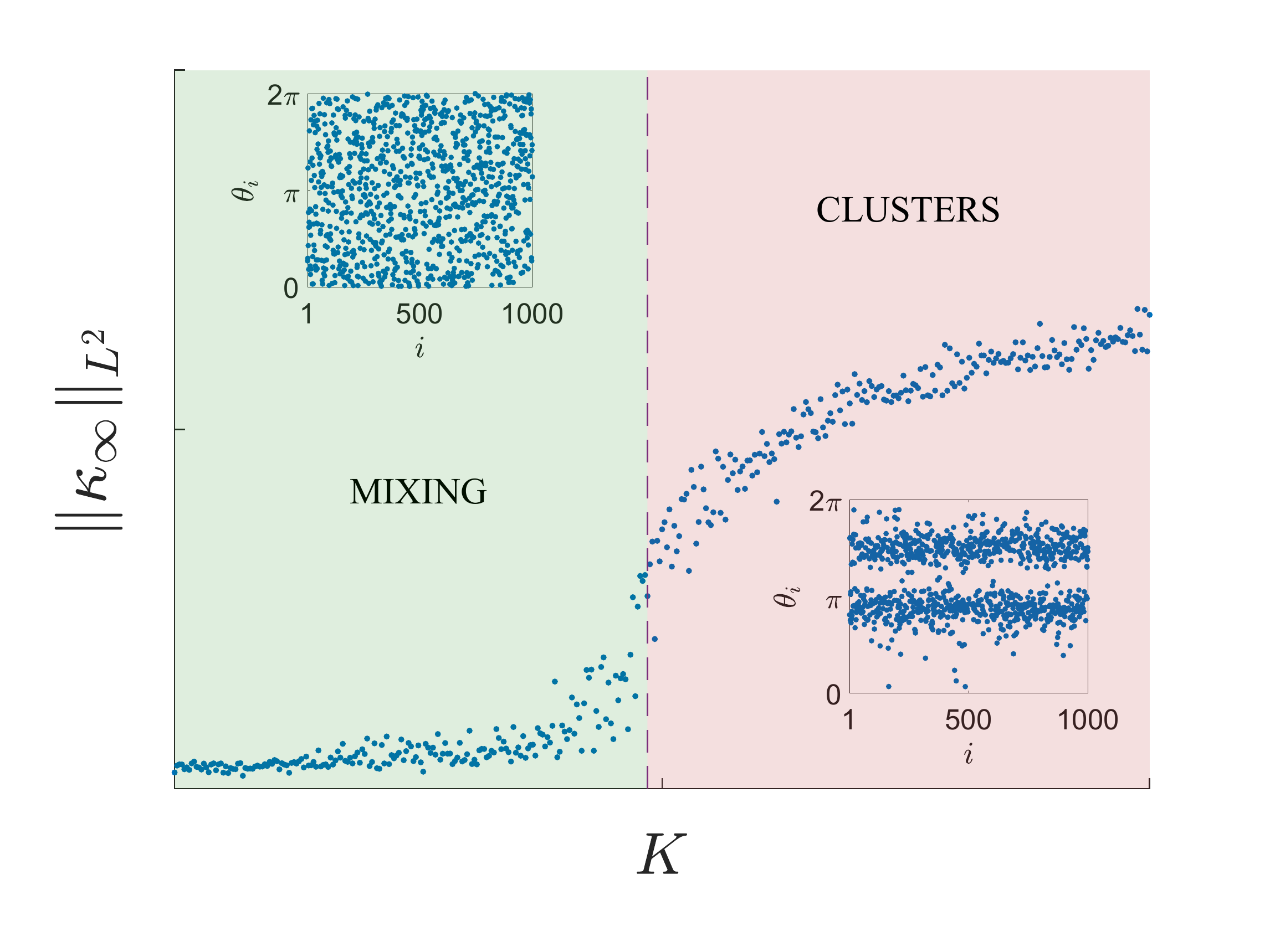}
	\caption{Andronov--Hopf bifurcation leading to the formation of two--cluster in the KM
          on complete graphs and bimodal $g$.}
	\label{f.b-aa}
      \end{figure}

\begin{figure}
	\centering
	\includegraphics[width = .6\textwidth]{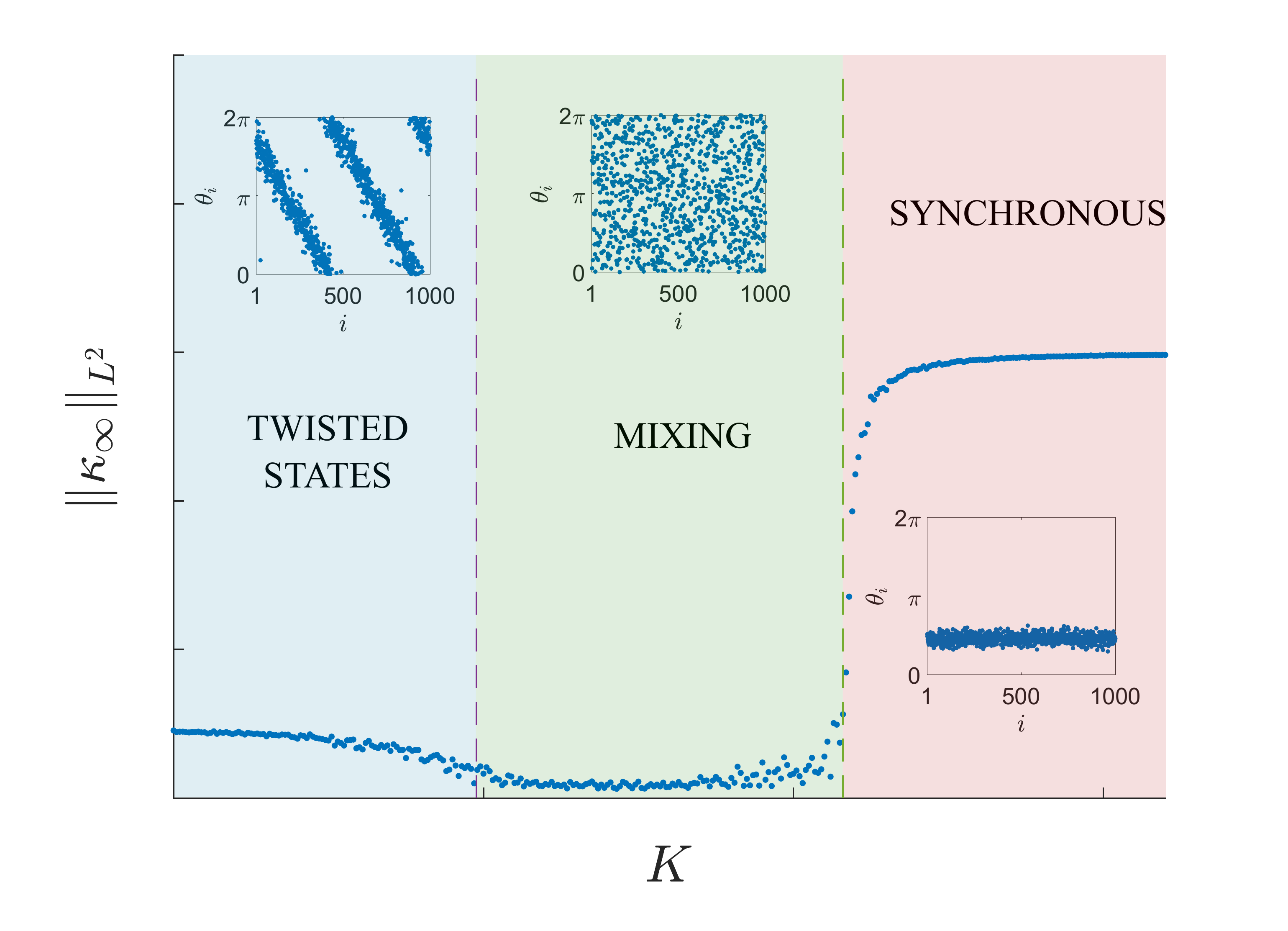}
	\caption{Bifurcations in the KM with nonlocal nearest--neighbor coupling
          and unimodal frequency distribution. The pitchfork bifurcation at $K_c^-<0$ leads to formation
         of a 2--twisted state. At $K_c^+>0$ the system undergoes a bifurcation leading to synchronization.}
	\label{f.u-nn}
\end{figure}

\begin{figure}
	\centering
	\includegraphics[width = .6\textwidth]{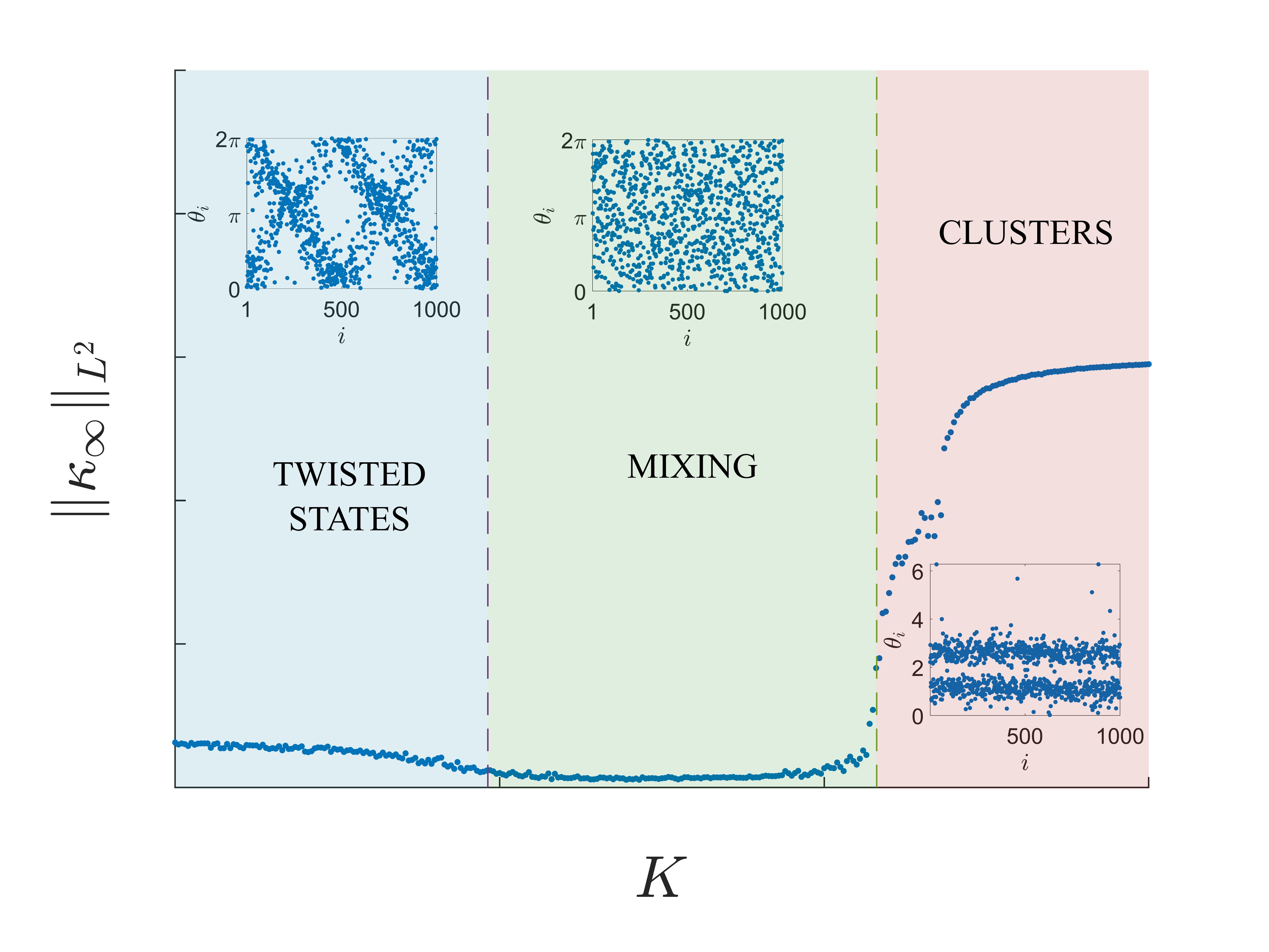}
	\caption{Bifurcations in the KM with nonlocal nearest--neighbor coupling
          and bimodal frequency distribution. The Andronov--Hopf bifurcation at $K_c^+>0$ leads to
          formation of 2--cluster, whereas the bifurcation at $K_c^-<0$
          results in  a pair of 2--twisted states traveling in opposite directions.
}
	\label{f.b-nn}
\end{figure}

Having understood the qualitative features of the critical curves for the unimodal and bimodal densities,
we now turn to bifurcations.
\begin{description}
\item[(Uaa)]
  We start with the all--to--all coupling first. In this case, $W\equiv 1$, the largest eigenvalue
of $\bW$ is $\mu=1,$ and the corresponding eigenfunction is constant \cite{ChiMed19a}.
Since the critical curve $G(\p\Pi)$ is bounded,  $2K^{-1}\notin \Pi$ for $K\gg 1$.
Thus, there are no roots of \eqref{spectral} for large $K$. As we decrease $K$,
$2K^{-1}$ hits $G(\p\Pi)$ when
$$
2K^{-1}=G(0)=\pi g(0).
$$
Note that the corresponding root of \eqref{spectral} is $z=0$. Thus, at $K_c=2/(\pi g(0))$
the system undergoes a pitchfork bifurcation. The emerging pattern is  determined by the
unstable mode $v_0$, which has a singularity at $\omega=0$\footnote{For $\phi$
  vanishing in a neighborhood of the origin $\langle  \Upsilon_0,\phi\rangle$ is a regular functional.}.
This implies that the emerging pattern contains a stationary cluster. Further, since
$V\equiv 1$, the bifurcating solution is uniform in space. We conclude that the instability
leads to the formation of a stationary coherent cluster (see Figure~\ref{f.u-aa}).
This is a classical scenario of the onset of synchronization.
\item[(Baa)]
  Next we discuss the case of the bimodal density and all--to--all coupling. In this case
  $2K^{-1}$ hits $G(\p\Pi)$ at the point of double intersection of the critical curve
  with the real axis:
  $$
  2K^{-1}=G(0)=\pi g(\pm y^\ast).
  $$
  The roots of \eqref{spectral} $z=\pm\1 y^\ast$. The system undergoes
  Andronov--Hopf bifurcation at $K_c=1/\left(\pi g(y^\ast)\right)$\footnote{The KM with all--to--all
    coupling and bimodal frequency distribution was discussed in \cite{Die16}, but the Andronov--Hopf
    bifurcation was not identified there.}
  The eigenfunctions
  $v_{\pm\1 y^\ast}$  have singularities at $\omega=\pm y^\ast$
  respectively (cf.~\eqref{viy-distribution}), while $V$ is still constant.
  Thus, the emerging pattern consists of two spatially homogeneous
  clusters rotating with constant speed in opposite directions
  (see Figure~\ref{f.b-aa}).
\item[(Unn)]
  It remains to consider the nonlocal nearest--neighbor coupling
 (see \cite[\S 5.2]{ChiMed19a} for the definition of $W$ in this case). 
 The new feature here is that along with the largest positive eigenvalue
 $\mu^+=1$ (corresponding to $V^+\equiv 1$), there  can be one or more
 negative eigenvalues (see \cite{ChiMed19a} for a detailed discussion).
 Suppose $\mu^-<0$ is the smallest negative eigenvalue of $\bW$.
 The corresponding eigenspace is spanned by $V^-_{1,2}=e^{\pm 2\pi\1 q x}$
 for some $q\in \N$. Thus, there are two bifurcation points
 $$
 K^+_c=2/(\pi\mu^+g(y^\ast))>0\quad\mbox{and}\quad
 K^-_c=2/(\pi\mu^-g(0))<0.
 $$
 $\Upsilon_0$ component of the unstable mode $v_0$ is the same for the bifurcations at $K_c^-$ and $K_c^+$.
 This implies that the bifurcating patterns gravitate towards stationary clusters.
 However, the spatial organization is different. The pattern emerging at $K_c^+$
 is uniform in space, whereas those emerging at $K_c^-$ are organized as
 $q$--twisted states (see Figure~\ref{f.u-nn}).
  
A new feature of this example is that in addition to positive eigenvalues of $\bW$ there are
negative eigenvalues. Denoting the largest postive and smallest negative eigenvalues of
$\bW$ by $\mu^+$ and $\mu^-<0$ respectively (see \cite{ChiMed19a} for explicit formulae
of the eigenvalues of $\bW$). For $\mu^+$ the corresponding eigenfunction is constant as
in the all--to--all coupling case. For $\mu^-$, the eigenfunctions are linear combinations
of so--called $q$--twisted states: $e^{\pm 2\pi\1 q x}$ for the appropriare $q\in \N$.
Thus, in the unimodal case the mixing state bifurcates into a spatially homogeneous solutions
at $K^+=2/(\pi\mu^+g(0))>0$ and into a twisted state at $K^-=2/(\pi\mu^-g(0))<0$
(see Figure~\ref{f.u-nn}). In the bimodal case, the mixing state bifurcates  into a two-cluster
at $K^+=2/(\pi\mu^+g(y^\ast))>0$ and into a pair of twisted states at $K^-=2/(\pi\mu^-g(y^\ast))<0$
(see Figure~\ref{f.u-nn}).
\item[(Bnn)]
  The only difference of this case with the one that we just discussed is that the principal unstable
  modes  $v_{\pm\1 y^\ast}$ are localized around $\omega=\pm y^\ast$. Thus, the stationary
    patterns in \textbf{(Unn)} turn into rotating ones: rotating clusters at $K_c^+$
    and twisted states traveling in opposite directions at $K_c^-$
(see Figure~\ref{f.b-nn}).
\end{description}

\begin{figure}
  \centering
  \includegraphics[width = .4\textwidth]{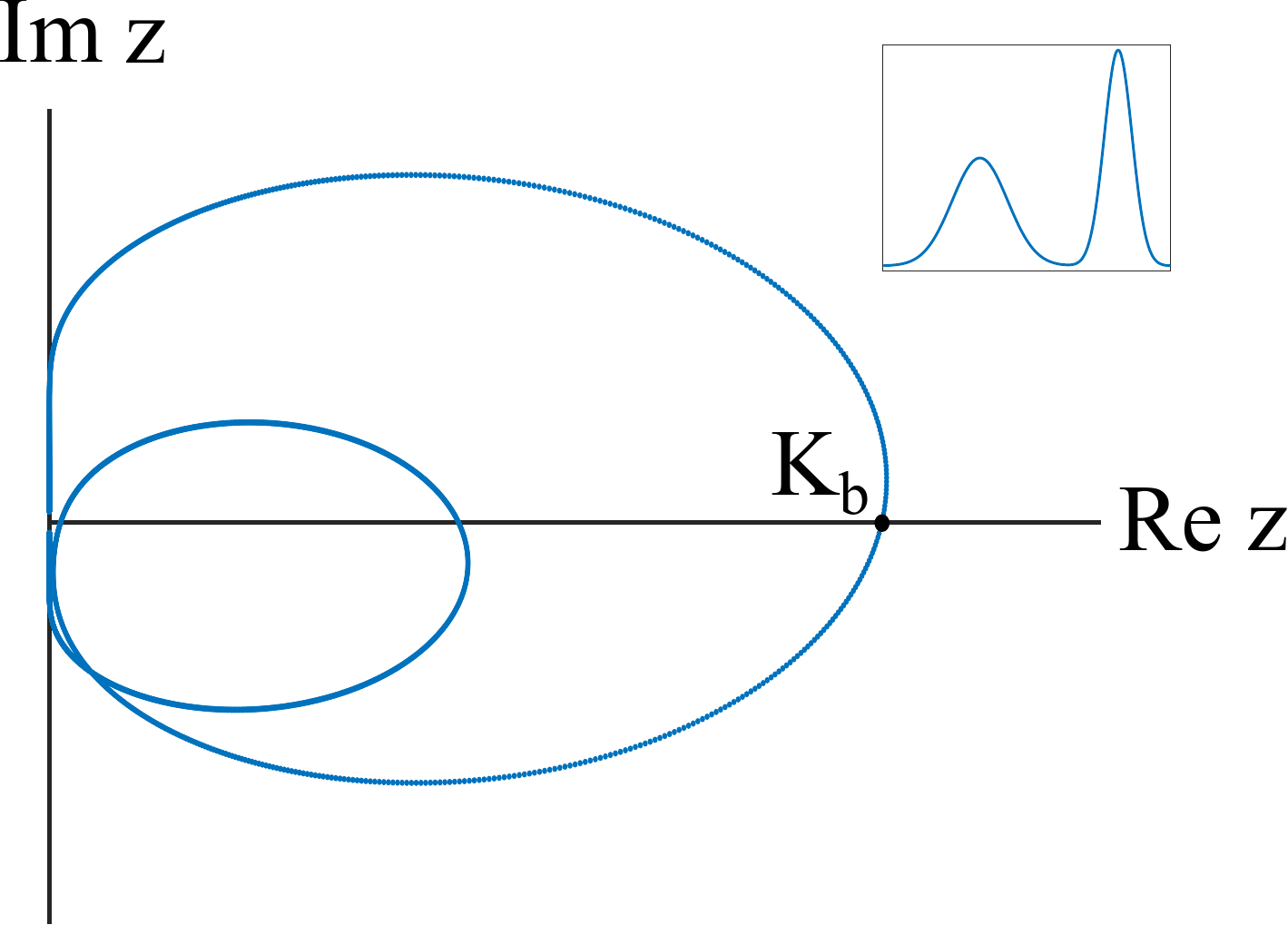}
	\caption{The critical curve for asymmetric bimodal distribution.}
	\label{f.critical-asym}
\end{figure}

\begin{figure}
  \centering
  \textbf{a}~\includegraphics[width = .4\textwidth]{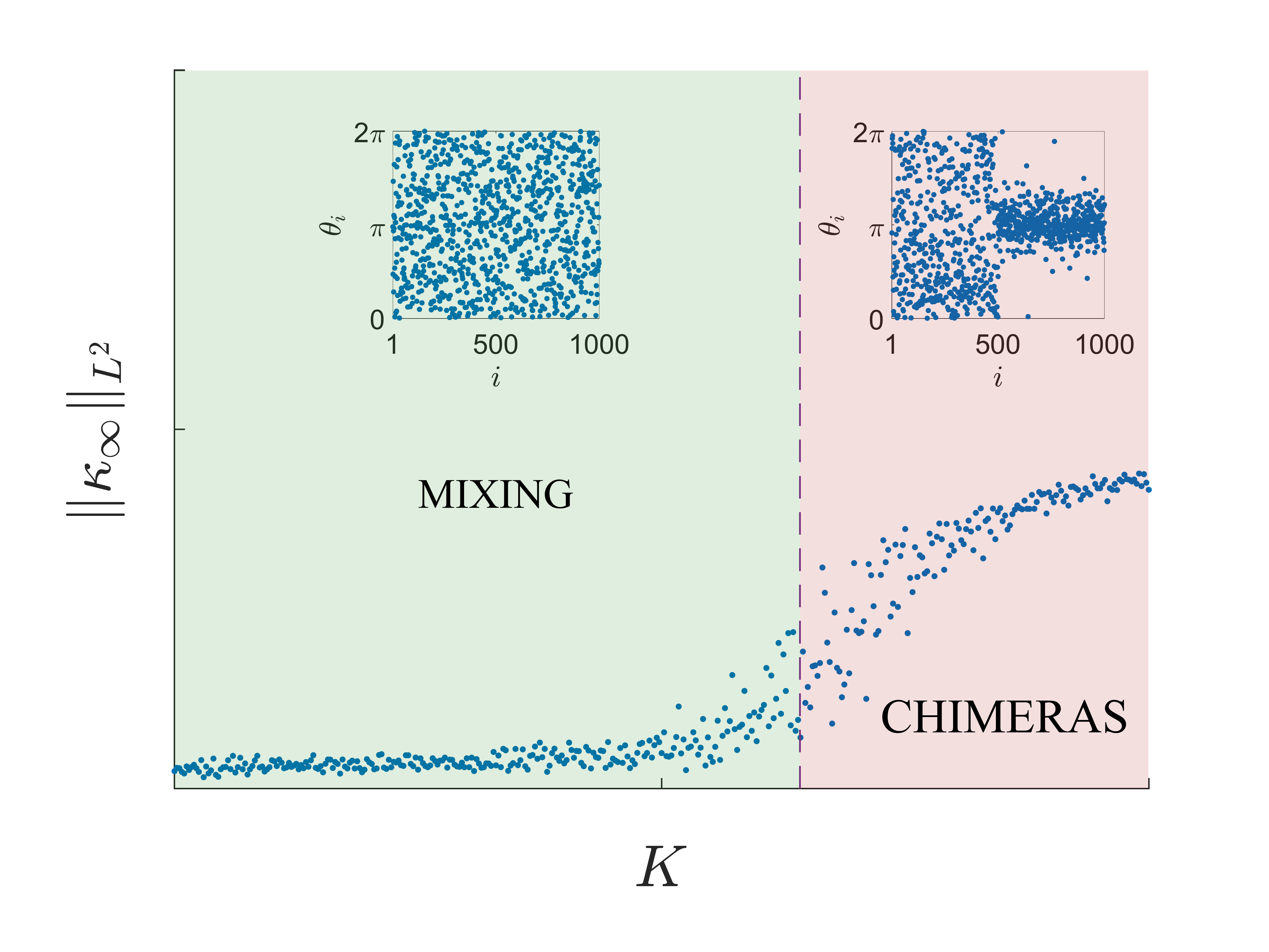}\quad\quad
  \textbf{b}~\includegraphics[width = .4\textwidth]{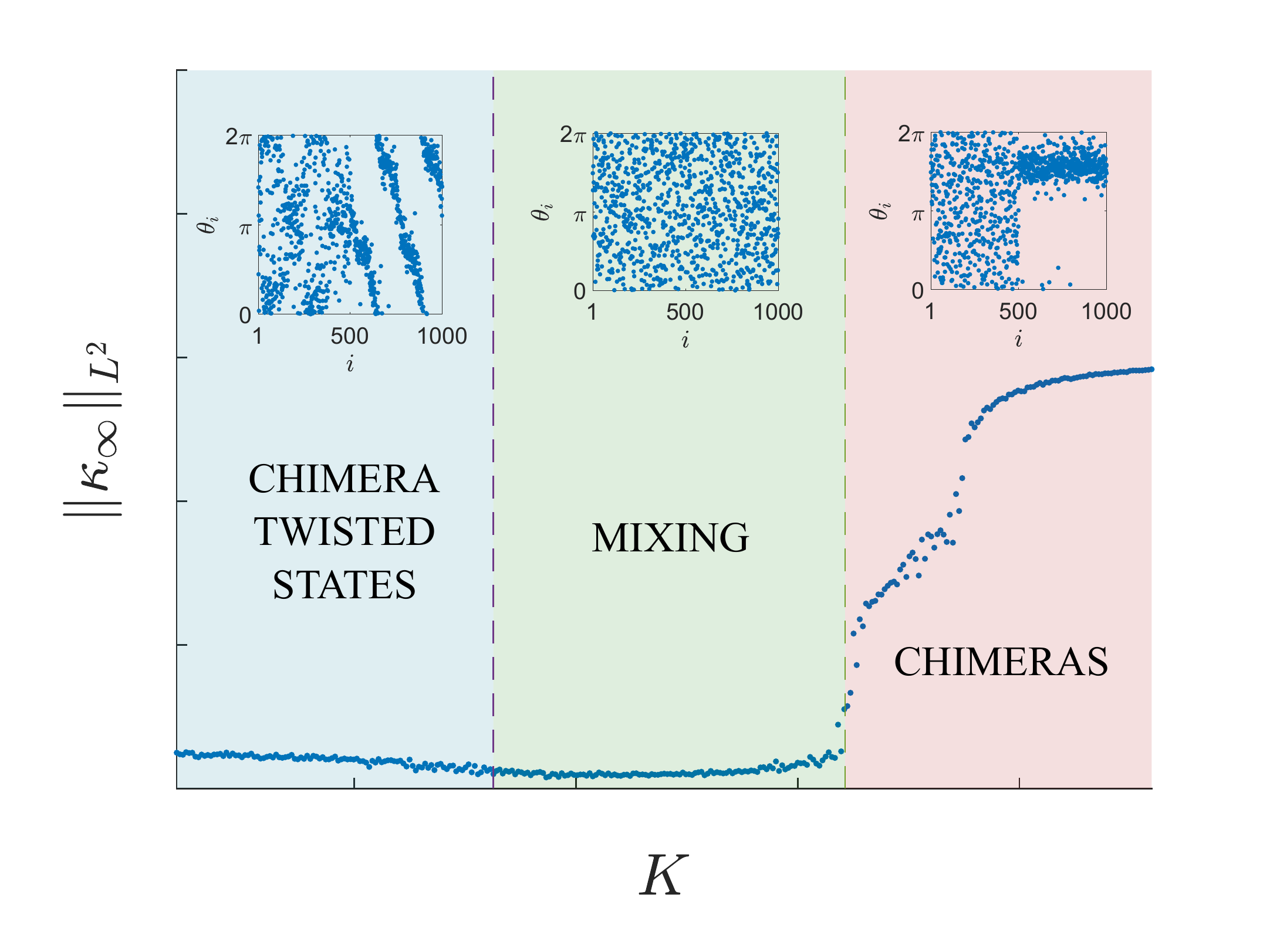}
  \caption{The bifurcation diagrams for the KM with asymmetric bimodal distribution
    (see Figure~\ref{f.critical-asym}). The coupling is all--to--all in \textbf{a} and nonlocal nearest--neighbor
    in \textbf{b}. To elucidate the structure of these patterns, we separated the oscillators into two groups
    depending on the sign of the intrinsic frequency.
}
	\label{f.asym-patterns}
\end{figure}

This concludes the description of the bifurcation scenarios in the KM with symmetric unimodal and bimodal
frequency distribution on complete and nonlocal nearest--neighbor graphs. Breaking the symmetry of the
distribution (see Figure~\ref{f.critical-asym}) results in new patterns including chimera like patterns shown in 
Figure~\ref{f.asym-patterns}. They can be understood using the techniques of this paper.
The situation is even more interesting for the second--order
KM, which will be covered in the future work \cite{CMM20+}. The
complete and nonlocal nearest--neighbor graphs were used in this paper as
representative examples. The analysis of this paper applies without
any changes to the KM
on a variety of convergent graph sequences including
Erd\H{o}s--R{\'e}nyi, small-world, and
power--law graphs (cf.~\cite{CMM18, Med19}). 


\vskip 0.2cm
\noindent
{\bf Acknowledgements.} This work was supported in part by NSF grants DMS 1715161 and 2009233 (to GSM).
Numerical simulations were completed using the high performance computing
cluster (ELSA) at the School of Science, The College of New Jersey. Funding of ELSA is provided in part by National Science
Foundation OAC-1828163. MSM was additionally endorsed by a Support of Scholarly
Activities Grant at The College of New Jersey.

\def\cprime{$'$} \def\cprime{$'$} \def\cprime{$'$}
\providecommand{\bysame}{\leavevmode\hbox to3em{\hrulefill}\thinspace}
\providecommand{\MR}{\relax\ifhmode\unskip\space\fi MR }
\providecommand{\MRhref}[2]{%
  \href{http://www.ams.org/mathscinet-getitem?mr=#1}{#2}
}
\providecommand{\href}[2]{#2}

\end{document}